%% file: main.tex
\documentclass[aps,twocolumn,reprint,showpacs,amsmath,amssymb,prl,superscriptaddress,floatfix]{revtex4-2}
\pdfoutput=1

\usepackage[pdftex]{graphicx}
\usepackage{bm}
\usepackage[pdftex,colorlinks=true,citecolor=blue,urlcolor=blue,linkcolor=black]{hyperref}
\usepackage{braket}
\usepackage[dvipsnames]{xcolor}
\usepackage{siunitx}
\usepackage{amsmath}
\usepackage{cancel}
\usepackage{soul}
\usepackage{mathrsfs}
\usepackage{comment}
\usepackage{lipsum} 
\usepackage{siunitx}
\usepackage[english]{babel}
\addto\captionsenglish{}
\usepackage{float}
\DeclareSIUnit{\lHO}{\ensuremath{\mathit{l}_{\mathrm{HO}}}}
\DeclareSIUnit{\pHO}{\ensuremath{\mathit{p}_{\mathrm{HO}}}}
\DeclareUnicodeCharacter{212B}{\AA}

\newcommand*\Methods{(Methods)}

\usepackage{float}

\usepackage{silence}
\begin{document}


\title{
Fermionic pairs -- from the surface to the bulk}
\author{Sandra Brandstetter}
    \email{brandstetter@physi.uni-heidelberg.de}
	\affiliation{Physikalisches Institut der Universit\"at Heidelberg, Im Neuenheimer Feld 226, 69120 Heidelberg, Germany}
    \affiliation{Current address: Department of Physics, Harvard University, Cambridge, Massachusetts 02138, USA}
\author{Carl Heintze}
	\affiliation{Physikalisches Institut der Universit\"at Heidelberg, Im Neuenheimer Feld 226, 69120 Heidelberg, Germany}
\author{Fabian Brauneis}
	\affiliation{Technische Universit\"at Darmstadt, Department of Physics, 64289 Darmstadt, Germany}
\author{Stephanie~M.~Reimann}
	\affiliation{Division of Mathematical Physics and NanoLund, LTH, Lund University, Box 118, SE-22100 Lund, Sweden}
\author{Georg M. Bruun}
	\affiliation{Department of Physics and Astronomy,
Aarhus University, Ny Munkegade 120, DK-8000 Aarhus C, Denmark}
\author{Maciej Ga\l ka}
	\affiliation{Physikalisches Institut der Universit\"at Heidelberg, Im Neuenheimer Feld 226, 69120 Heidelberg, Germany}
\author{Selim Jochim}
	\affiliation{Physikalisches Institut der Universit\"at Heidelberg, Im Neuenheimer Feld 226, 69120 Heidelberg, Germany}

\date{\today}

\begin{abstract}
 
Fermion pairing underlies collective quantum phenomena across widely different forms of matter. In extended systems such as  ultracold Fermi gases, pairing is commonly understood through the BCS--BEC crossover, where the pair size evolves from large, overlapping Cooper pairs to tightly bound dimers. In finite systems such as atomic nuclei, superconducting grains and quantum dots, however, the same pairing tendency competes with confinement, shell filling and spatial inhomogeneity, making the microscopic structure of pairs much harder to access.  Here, we image pair correlations in a finite, tunable system of few fermionic atoms with single-particle resolution and full counting statistics. We observe that confinement and shell structure re-organize pairing in real space: In the weakly interacting, confinement-dominated regime, closed-shell configurations suppress correlations in the high-density trap center. Pairing is mainly observed toward the low-density surface. Open-shell systems, however, support substantially stronger central pairing. Already for surprisingly small systems, increasing either interaction strength or particle number restores a locally bulk-like Cooper-pair profile in the trap center, whereas the edge retains dimer-like correlations.  By resolving where pairs form and how their character changes from localized dimers to overlapping Cooper pairs, our measurements provide a microscopic view of pairing in finite fermionic matter and connect the physics of mesoscopic cold atoms to pairing phenomena in nuclei and superconducting nanostructures.

\end{abstract}

\maketitle



\begin{figure}
\centering\includegraphics{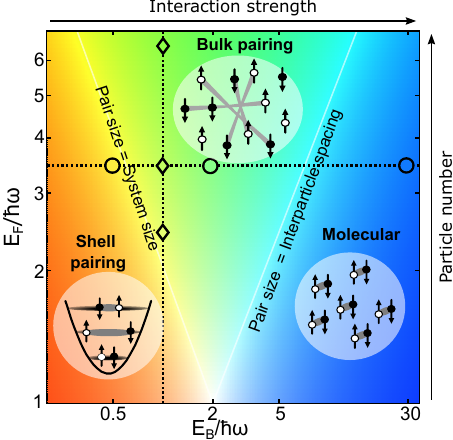}
    \caption{\textbf{Pairing regimes.} The different pairing regimes determined by the
    interplay of the interaction strength and particle number, or equivalently, the pair size $r_{\mathrm{B}}$, system size $r_{\mathrm{F}}$, and mean interparticle spacing $\ell$.   The plot encodes the competition of length scales at the trap center in an RGB color map, $(R,G,B) =(1/r_{\mathrm{F}},\,1/\ell,\,1/r_{\mathrm{B}})$, normalized to its largest component. The boundaries indicate crossovers between dominant length-scale hierarchies rather than sharp phase transitions . Confinement-dominated shell pairing occurs for $r_{\mathrm{B}} \gtrsim r_{\mathrm{F}}$ (red). Here, $r_{\mathrm{B}}$ should be understood as the two-body interaction length scale, while the physical pair extent is limited by the trap. Bulk  pairing for $\ell <  r_{\mathrm{B}} \leq r_{\mathrm{F}}$ (green), and molecular dimers are formed
    for $r_{\mathrm{B}} < \ell  \leq r_{\mathrm{F}}$ (blue).  Open circles and diamonds denote the parameters used in Figs.~2 and~3.}
    \label{fig: Pairing regimes}
\end{figure}

Pair formation is a central feature of interacting fermionic matter, and its associated correlations provide a powerful probe of systems ranging from nuclei and superconductors to ultracold atomic gases. In extended systems, the character of pairing is governed by the relation between the pair size and the mean interparticle spacing. Tightly bound dimers give rise to a Bose--Einstein condensate (BEC), whereas large, strongly overlapping Cooper pairs~\cite{Cooper_1956, Bardeen_1957} define the Bardeen--Cooper--Schrieffer (BCS) regime. The smooth crossover between these limits has become a central paradigm for fermionic superfluidity. Ultracold atomic gases have played a decisive role in exploring this BCS--BEC crossover~\cite{Regal_2004, Greiner_2005, Jochim2003, Zwierlein_2004, Greiner2003, Zwierlein2005}, owing to their exceptional tunability of interaction strength, density, and geometry. More recently, this control has enabled precision probes of correlations in quasi-infinite systems with single-particle resolution~\cite{Daix2026,Yao2025,deJongh2025}.

In finite fermion systems, however, pairing manifests in fundamentally different ways than in the homogeneous limit. This is most prominently known from nuclear physics~\cite{Bohr1958,BrinkBroglia, FiftyYearsBCS} and has theoretically also been discussed for few-body cold-atom systems~\cite{Heiselberg_2002,Heiselberg_2003,Bruun_2002,Bruun_2002b,Rontani_2009}. When the pair size becomes comparable to  the system itself, pairing enters a confinement-dominated regime in which correlations are shaped not only by interactions but also by confinement, which imposes a discrete level structure and a finite spatial extent.  In this limit, finite-size systems also exhibit closed-shell configurations for so-called "magic" particle numbers in which pairing is suppressed, because the density-of-states at the Fermi energy vanishes. 

Here, we experimentally investigate fermionic pairing in the finite-size regime, using single-particle-resolved measurements to reveal how confinement reshapes correlations. By independently tuning the interaction strength and particle number, we control the relevant length scales of the problem: the mean interparticle spacing $\ell$, the two-body length scale $r_{\mathrm{B}}$ associated with the bound-state energy, and the overall system size $r_{\mathrm{F}}$. In a trap, $r_{\mathrm{B}}$ can formally exceed $r_{\mathrm{F}}$; in this regime, confinement limits the physical extent of the pair and strongly shapes pairing through the discrete level structure. This tunability allows us to access not only the pairing regimes familiar from the homogeneous limit, where the pair size is smaller than the system size, but also the confinement-dominated regime in which pairing is governed by the trap itself.

We show how the interplay between the length scales  $r_{\mathrm{B}}$, $\ell$, and $r_{\mathrm{F}}$ gives rise to three distinct pairing regimes, illustrated in Fig.~\ref{fig: Pairing regimes}. This length-scale competition is the natural way in which the BCS-to-BEC crossover is realized in two dimensions, where there is no unitary point: the crossover is instead governed by the ratio of the pair binding energy to the local Fermi energy. When both $r_{\mathrm{B}}$ and $\ell$ are much smaller than $r_{\mathrm{F}}$, the gas behaves as locally homogeneous. This yields a Cooper-pair regime of delocalized pairs in real space when $\ell < r_{\mathrm{B}}  \leq r_{\mathrm{F}}$. The crossover to the molecular regime of tightly bound pairs is realized as the density decreases or the binding energy increases such that ($r_{\mathrm{B}} < \ell  \leq r_{\mathrm{F}}$).



In the opposite limit where $r_{\mathrm{B}} \gtrsim r_{\mathrm{F}}$, pair formation is strongly influenced by confinement and the discrete level structure that is imposed by it. Thus, pairing cannot be inferred from local quantities.


\begin{figure*}
    \includegraphics{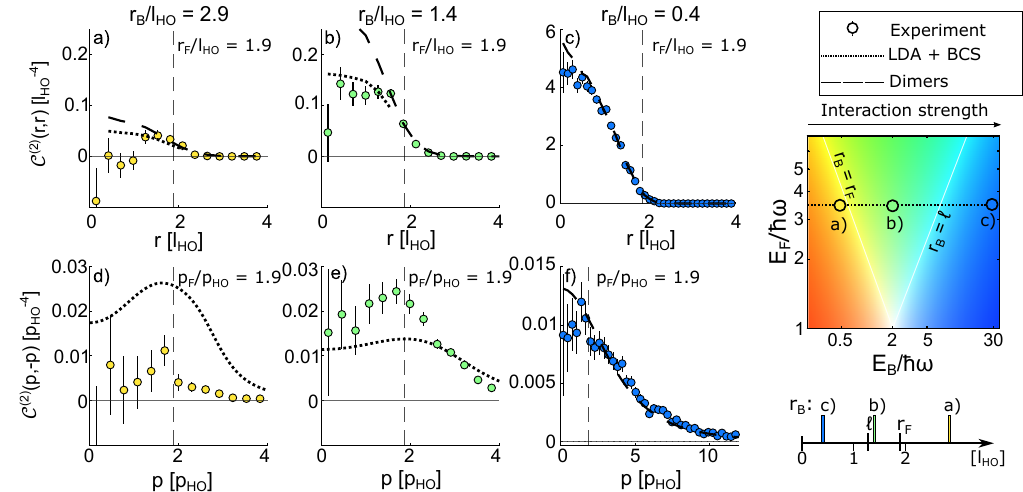}
    \caption{\textbf{Crossing pairing regimes.}
    Pair correlations in real (a-c) and momentum (d-f) space of $6+6$ atoms prepared in the ground state of a 2D harmonic oscillator potential at different interaction strengths.
    \textbf{(a-c),}  The real space  function  $\mathscr{C}_2(r,r)$ giving the correlations between 
    spin $\uparrow$ and spin $\downarrow$ atoms to be at the same position at radius $r$, within the experimental resolution. \textbf{(d-f)} 
    The momentum space  function  $\mathscr{C}_2(p,-p)$ giving the 
    correlations between 
    spin $\uparrow$ and spin $\downarrow$ atoms to have opposite momenta with magnitude $p$. The color of the data points reflects the pairing regime according to Fig 1. The measured data points are compared to the pair correlation function calculated within BCS theory assuming bulk 
    $({\mathbf p},\uparrow)\leftrightarrow (-{\mathbf p},\downarrow)$ pairing combined with LDA (dotted curve) and assuming 
    a gas of tightly bound dimers (dashed-dotted curve). The dashed vertical lines mark the Fermi radius (a-c) or Fermi momentum (d-f) obtained from the non-interacting ground state configuration. All error bars represent the standard error of the mean.}
    \label{fig: Interactions}
\end{figure*}

\section{Experimental setup and observables}

We start our experiments by preparing a balanced mixture of $^6$Li atoms in two hyperfine states (denoted as spins $\sigma=\{\uparrow,\downarrow$\}) in the ground state of an optical tweezer. Along the transverse direction, a vertical lattice provides tight confinement, rendering the system quasi two-dimensional (2D). Following Refs.~\cite{Bayha_2020, Serwane_2011}, we precisely control the fermion number $N$ per spin state. In a 2D harmonic oscillator, complete filling of the energy levels with energy $E_\text{n} =(n+1)\hbar\omega_\text{r}$ and degeneracy $n+1$ up to the Fermi level $n_\text{F}$ yields $N = (n_\text{F}+1)(n_\text{F}+2)/2$. The Fermi energy is given by $E_\text{F} = (n_\text{F} +3/2) \hbar \omega_\text{r}$ when the last shell is completely filled and   $E_\text{F} = (n_\text{F} +1) \hbar \omega_\text{r}$ when the last shell is not fully occupied, and sets the system radius $r_\text{F} = \sqrt{E_\text{F}/m_a\omega_r^2}$  and the mean interparticle spacing $\ell = \sqrt{2\pi\hbar^2/m_\text{a}E_\text{F}}$. Here, $m_\text{a}$ is the atomic mass of $^6$Li  and $E_\text{F}$, $r_\text{F},\ \ell$ are all defined assuming non-interacting particles. We control the interaction strength using a magnetic Feshbach resonance~\cite{Zuern_2013} and express it through the energy of the two-body (dimer) bound  state $E_\text{B}$.  We obtain the two-body binding energy in the trap following~\cite{Idziaszek_2006}. From $E_\text{B}$, we define the characteristic pairing length scale  $r_\text{B} = 2\sqrt{\hbar^2/(m E_\text{B})}$. Note that in extended 2D system a bound state exists at any interaction strength. It can formally exceed the system radius ($r_\text{F}$). In that regime, however, the spatial extent of the pairs is limited by the confining potential, which affects the structure and formation of pairs. \Methods

To probe pairing across these regimes, we image the atoms in real~\cite{Brandstetter_2024} and momentum space~\cite{Holten_2022} with single atom and spin resolution~\cite{Bergschneider_2018,Holten_2022}, and extract the density-density correlation function for atoms with opposite spin 
\begin{equation}
  \mathscr{C}^{(2)}(\mathbf{r}_\uparrow,\mathbf{r}_\downarrow) = \braket{\hat{n}_\uparrow(\mathbf{r}_\uparrow) \hat{n}_\downarrow(\mathbf{r}_\downarrow)}- \braket{\hat{n}_\uparrow(\mathbf{r}_\uparrow)}\braket{\hat{n}_\downarrow(\mathbf{r}_\downarrow)}.
	\label{eq: pairing_general_corr}
\end{equation}
from roughly 1000 repeated measurements. Here, $\mathbf{r}_\sigma$ denotes a point in 2D space, $\hat{n}_\sigma$ is the density operator for spin $\sigma$, and $\braket{...}$ denotes the expectation value.  We obtain the real space pair correlation function $\mathscr{C}^{(2)}(r,r)$ by counting those pairs of spin up and spin down atoms that are at the same position $\mathbf{r}_\uparrow=\mathbf{r}_\downarrow$ at a given radius $r=|{\mathbf r}_\sigma|$ within our resolution ($\delta r_\text{res} = \SI{300}{\nano \meter}$). \Methods

Single-atom-resolved measurements provide direct access to where pair correlations emerge within the trap. Although our resolution does not allow us to extract the contact from short-range correlations~\cite{Daix2026}, and the finite size of the system limits the applicability of bulk BCS-based inference~\cite{Yao2025,Obeso-Jureidini2022}, our real-space observable directly probes pairing in a regime where confinement, shell structure and spatial inhomogeneity are essential.

Analogously, we extract the momentum space pair correlations $\mathscr{C}^{(2)}(p,-p)$ from $\mathscr{C}^{(2)}(\mathbf{p}_\uparrow,\mathbf{p}_\downarrow)$ by  considering those atoms that sit at opposite momenta -- within our resolution $\delta p_\text{res} = \SI{0.06}{\pHO}$. Here and in the following, we give the momenta and positions in units of the harmonic oscillator momentum ($p_\text{HO} = \sqrt{\hbar m_a \omega_\text{r}}$) and length ($l_\text{HO} = \sqrt{\hbar/m_a \omega_\text{r}}$), respectively.

\section{Evolution of pairing with interaction strength}
The evolution of pairing across the different regimes is revealed when we follow a closed-shell system of $6{+}6$ atoms -- corresponding to the lowest three shells filled in a non-interacting system -- across different interaction strengths, i.e. different values of $r_{\mathrm{B}}$, crossing the diagram of Fig.~1 along the horizontal line. Here, the system size is $r_{\mathrm{F}} \approx \SI{1.87}{\lHO}$ and the mean interparticle spacing is $\ell \approx \SI{1.34}{\lHO}$.
In Fig.~2 we show the measured pair correlations in real (a-c) and momentum space (d-f) revealing how pair correlations evolve across the three different regimes.

First, we consider the molecular regime (Fig.~2c  and f) where $E_{\mathrm{B}}/\hbar\omega_{\mathrm{r}} = 30$ and $r_{\mathrm{B}} = \SI{0.37}{\lHO}$ so that the dimer size is the smallest length scale in the system, satisfying $r_{\mathrm{B}} \ll \ell \leq r_{\mathrm{F}}$. In this limit, the system consists to a good approximation of tightly bound dimers forming all over the system as seen in Fig.~\ref{fig: Interactions}c. The correlator $\mathscr{C}^{(2)}(r,r)$ is in this limit expected to be proportional to the density 
with a logarithmic divergence cut-off by  finite range effects and experimental resolution \Methods. Indeed, fitting $\mathscr{C}^{(2)}(r,r)$ with the density (dashed line) gives excellent agreement throughout most of the trap. 
Deviations appear only at the trap center, where the density is the highest and the mean interparticle spacing is  smallest so that  the dimers start to overlap.

The momentum distribution of a system of well-separated dimers is dominated by the two-body dimer wavefunction, i.e.
$\mathscr{C}^{(2)}(p,-p)\propto|\tilde\phi(p)|^2$ where $\tilde\phi(p)$ is the momentum-space dimer wave function~\cite{Idziaszek_2006}. In Fig.~\ref{fig: Interactions}f we show $\mathscr{C}^{(2)}(p,-p)$ and find that pair correlations extend over a broad range of momenta and are in  very good agreement with this theoretical prediction.

Next, we enter the intermediate regime shown in Fig.~\ref{fig: Interactions}b, where $E_{\mathrm{B}}/\hbar\omega_{\mathrm{r}}=2$ and $\ell < r_{\mathrm{B}}=\SI{1.4}{\lHO} < r_{\mathrm{F}}$. In this regime, the radial density gradient exposes different pairing behavior within the same trapped cloud. Near the low-density edge around $r_{\mathrm{F}}$, $\mathscr{C}^{(2)}(r,r)$ remains proportional to the density, consistent with isolated dimer-like pairs. Toward the trap center, where the density increases, the correlator instead saturates, indicating the onset of spatial overlap between pairs.

We compare this central-region behavior with a local-density BCS prediction that assumes bulk-like Cooper pairing between $({\mathbf p},\uparrow)$ and $(-{\mathbf p},\downarrow)$ states. The pairing gap is evaluated locally as $\Delta(r)=\sqrt{2E_{\mathrm{F}}(r)E_{\mathrm{B}}}$~\cite{Randeria_1989}, with $E_{\mathrm{F}}(r)=\pi\hbar^2 n(r)/m$ set by the total density $n(r)$ of both spin components. The corresponding short-distance real-space correlator, regularized by the finite interaction range and experimental resolution, is described in the Methods section. The theoretical curves are restricted to radii where $\ln(k_{\mathrm{F}}a_{\mathrm{2D}})<0$, for which the weak-coupling BCS expression is applicable. The resulting prediction agrees well with the measured correlations at the trap center, suggesting that bulk-like Cooper pairing emerges locally already for twelve atoms.

This interpretation is supported by the momentum-space correlations shown in Fig.~\ref{fig: Interactions}e. In contrast to the broad two-body momentum distribution observed for tightly bound dimers, $\mathscr{C}^{(2)}(p,-p)$ now peaks near the Fermi momentum, as expected for overlapping Cooper pairs. We find a surprising qualitative agreement by assuming locally homogeneous BCS-correlations $\mathscr{C}^{(2)}(p,-p)\propto\Delta^2/(\xi_p^2+\Delta^2)$, with $\xi_p=p^2/2m-E_{\mathrm{F}}$, and integrating over the density, restricted to radii where $\ln(k_{\mathrm{F}}a_{\mathrm{2D}})<0$.


To explore the confinement-dominated  regime, we reduce the attraction further to $E_\text{B} = 0.47 \hbar\omega_\text{r}$, such that $r_\text{B} = \SI{2.8}{\lHO}>r_\text{F}>\ell$. 
Here, the  confining potential strongly influences pairing, resulting in a spatial structure significantly different from the other pairing regimes. In Fig.~\ref{fig: Interactions}a,d we show the measured pair correlations in this regime in real and momentum space, respectively. We observe that pair formation is strongly suppressed in the trap center following neither the theoretical predictions in the BCS nor in the dimer regime, and it increases towards the edge of the trap peaking roughly around the Fermi radius. At the surface the observed pair density however surprisingly again follows the theoretical prediction for isolated dimers.



In this weakly-interacting regime where $r_B$ is  larger than $r_\text{F}$,  the pairing gap predicted from bulk BCS theory $\Delta = \sqrt{2 E_{\mathrm{F}} E_{\mathrm{B}}}$
becomes smaller than the trap level spacing $\hbar \omega$. Here, the system enters a shell pairing regime where the pairs form between time-reversed states within each harmonic oscillator shell, with the same principal quantum number $n$ but opposite angular momentum quantum number $m$~\cite{Heiselberg_2002,Heiselberg_2003,Bruun_2002,Bruun_2002b}, in close analogy to pairing between time-reversed states in dirty superconductors~\cite{Anderson1959}. Interestingly, the shell pairing regime realized here is similar to the case in atomic nuclei where the pair energy is typically much smaller than the level spacing. Indeed an enhancement of pairs near the surface has also been predicted for nuclei~\cite{Giovanardi2002,Sandulescu2005,Pillet2007}.

There are two distinct cases of shell pairing:
the "magic number" closed shell configuration where all harmonic oscillator shells up to the Fermi level are completely filled in the non-interacting limit, and the open-shell case where the highest occupied level is only partially filled. 
Within mean-field BCS theory, pairing between time reversed states occurs at arbitrarily low binding energies for the open shell case. However, Pauli blocking suppresses pairing in the closed shell case until the interaction is strong enough that it becomes energetically favorable to promote particles from the highest filled level to the lowest unoccupied one, giving rise to a quantum phase transition ~\cite{Bruun_2014}. In the few-body limit, this phase transition can be resolved as a cross-over~\cite{Bjerlin_2016} as was recently observed spectroscopically~\cite{Bayha_2020}.

\begin{figure}
    \centering \includegraphics[width=\columnwidth]{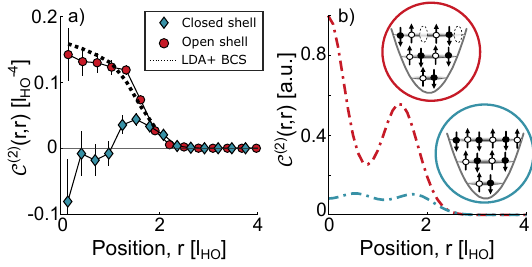}
    \caption{\textbf{Open and closed shell pairing} a) Measured real space pair correlations for a system with a partially filled  (5+5 atoms, red circles)  and completely filled (6+6 atoms, blue diamonds) highest harmonic oscillator shell at a constant interaction strength $E_\text{B}/\hbar\omega_\text{r} = 0.47$. The dotted curve shows the pair correlation function calculated assuming bulk  $({\mathbf p},\uparrow)\leftrightarrow (-{\mathbf p},\downarrow)$ Cooper pairing. Figure b) shows the comparison to the pair correlation function calculated assuming $(n,m,\uparrow)\leftrightarrow(n,-m,\downarrow)$ pairing in both regimes.  All error bars represent the standard error of the mean.
    }
    \label{fig: Open_at_the_close}
\end{figure}

To isolate the role of shell filling,  we compare the real-space pair correlations for systems of $6{+}6$ atoms corresponding to three completely filled shells to the open shell system of $5{+}5$ atoms, where the last filled shell has two holes.  The measured correlations are shown in Fig.~\ref{fig: Open_at_the_close}a. In both cases, the two-body binding energy is $E_{\mathrm{B}}/\hbar\omega_{r}=0.47$,
i.e.\ the same as in Fig.~\ref{fig: Interactions}a. This interaction strength lies below the critical binding energy  for the closed-shell case~\cite{Bayha_2020}. Notably, we observe that this leads to a marked suppression of pairing in the bulk part of the cloud so that the pairing is maximal near the surface, whereas there is no such suppression for the open shell case where pairing is maximal in the center where the density is largest. 

Following Ref.~\cite{Bruun_2014}, we compare our measurements with theoretical calculations, solving the BCS equations in the shell-pairing regime, i.e. restricting pairing to particles in time-reversed states with quantum numbers $(n,m,\uparrow)$ and $(n,-m,\downarrow)$, \Methods. For the open-shell system, the pairing model includes all shells up to the Fermi surface, whereas for the closed-shell system it includes all levels up to the first unoccupied shell. The resulting correlation functions are shown in Fig.~\ref{fig: Open_at_the_close}b. The calculation predicts substantially stronger pairing in the open-shell system and a corresponding suppression in the closed-shell system, in agreement with the experimental data. An LDA mean-field BCS description based on bulk-like Cooper pairing between $(\mathbf{p},\uparrow)$ and $(-\mathbf{p},\downarrow)$ states captures the open-shell data quantitatively, but fails for the closed-shell system at the same interaction strength. The resulting discrepancy identifies the suppressed central pairing as a genuine shell-filling effect beyond a local bulk description.


\section{Evolution of pairing with particle number}

\begin{figure}
    \includegraphics{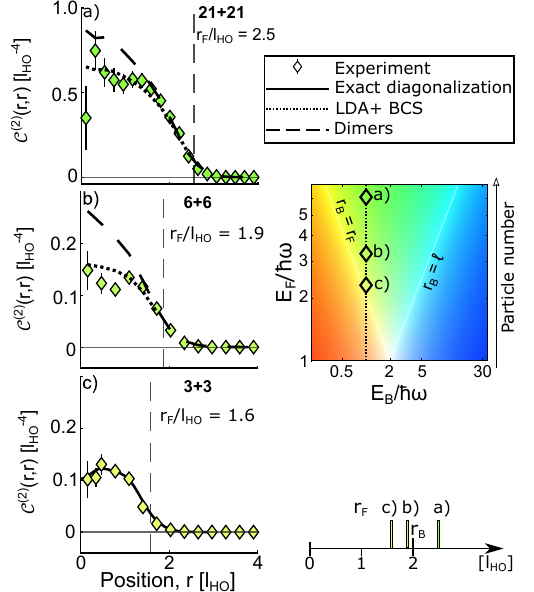}
    \caption{\textbf{From few-body physics to bulk Cooper pairing} Measured real space pair correlations for different atom numbers at a constant binding energy $E_\text{B}/\hbar\omega_{\mathrm{r}} = 1 $. The color of the data points reflects the pairing regime according to Fig 1, as shown in the inset. The results for the smallest system (3+3 atoms) are compared to the pair correlations obtained from exact diagonalization of the many body Hamiltonian (solid curve). The dotted curve gives $\mathscr{C}^{(2)}(r,r)$ 
    calculated assuming bulk $({\mathbf p},\uparrow)\leftrightarrow (-{\mathbf p},\downarrow)$ Cooper pairing, and the dashed curve shows 
    $\mathscr{C}^{(2)}(r,r)$ assuming a gas of dimers.  The dashed vertical line marks the Fermi radius obtained from the non-interacting ground state configuration. All error bars represent the standard error of the mean.
    }
    \label{fig: Particle number}
\end{figure}

Moving along the vertical axis of the pairing diagram, i.e., varying the particle number at fixed interaction strength, reshapes the hierarchy of length scales, tuning the system from the shell pairing to the bulk pairing regime. What distinguishes this trajectory, however, is that it simultaneously carries the system from a few-body regime accessible to exact diagonalization into one where mean-field and other many-body descriptions are expected to become applicable. Figure \ref{fig: Particle number} shows the measured pair correlations in real space for a system with a binding energy of $E_\text{B}/\hbar \omega_\text{r} = 1$ i.e.  $r_\text{B} = \SI{2}{\lHO}$ and different particle numbers. 

In the few-body closed shell configuration  of $3+3$ atoms ($E_{\mathrm{F}}/\hbar\omega_{\mathrm{r}} = 2.5$, $r_{\mathrm{F}} = \SI{1.55}{\lHO}$) shown in Fig.~\ref{fig: Particle number}c, we observe the characteristic suppression of shell pairing  at the trap center.  We compare the experimental results to the real space pair correlation function obtained with exact diagonalization \Methods, which is possible in this few-body limit~\cite{Bjerlin_2016,Rontani_2009,Rontani2017}. There is excellent quantitative agreement between the measured and calculated correlations, which provides a  benchmark of our experimental results. 

For $6{+}6$ atoms ($E_{\mathrm{F}}/\hbar\omega_{\mathrm{r}} = 3.5$, $r_{\mathrm{F}} = \SI{1.87}{\lHO}$, also shown in Fig.~\ref{fig: Particle number}b), the pair size becomes comparable to the system size and the system enters the crossover between the shell and bulk pairing regimes. Here, pairing is already remarkably well described by bulk BCS theory in the center and evolves into dimer-like behavior toward the low-density edge. For $21{+}21$ atoms ($E_{\mathrm{F}}/\hbar\omega_{\mathrm{r}} = 6.5$, $r_{\mathrm{F}} = \SI{2.55}{\lHO}$), the system size exceeds the pair size, and the measured pair distribution in the trap center agrees closely with the bulk BCS predictions, again evolving into dimer-like behavior as the density decreases.

\section{Conclusion and Outlook}

Our measurements show that in the weakly interacting limit, confinement and shell filling do not merely change the strength of fermionic pairing, but reorganize its spatial structure. In closed-shell configurations, pair correlations are suppressed in the high-density trap center and enhanced near the surface, whereas open-shell systems support central pairing. 

With increasing interaction strength or particle number, this shell-dominated profile evolves into locally bulk-like Cooper-pair correlations in the trap center, while correlations at the low-density edge are consistent with tightly bound dimers. This behavior resembles that of macroscopic two-dimensional systems, where the BCS-to-BEC crossover is realized as the density, and thus the local Fermi energy, drops below the pair binding energy \cite{Murthy_2017}. It is remarkable to observe such crossover-like behavior in few-body systems, where a clear separation of scales is not established.

 Such robustness of the LDA in a finite, inhomogeneous paired system is not unique to trapped atomic gases: similar behavior has been reported in nuclear density-functional calculations, where pairing fields inferred from uniform nuclear matter reproduce finite-nucleus Hartree-Fock-Bogoliubov pairing fields with unexpected accuracy~\cite{margueron_effective_2008, kucharek_pairing_1989}. 

The experiment thereby establishes a spatially resolved quantum simulator of finite paired matter, bridging exactly solvable few-body systems and locally homogeneous many-body regimes. Looking ahead, measurements that resolve orbital quantum numbers $(n,m)$, together with controlled trap deformations, spin imbalance, or interaction quenches, could probe pairing selection rules, collective pairing modes, and surface-pairing dynamics in finite fermionic systems.

\bibliography{Realspace}

\paragraph*{Competing Interest}
The authors declare no competing interests.

\paragraph*{Correspondence and requests for materials}
should be addressed to S.B.

\paragraph*{Acknowledgements}
S.M.R. and F.B. thank J. Bjerlin for input and numerical advice at an early stage of the project. 

\paragraph*{Funding}
The work was financially supported by DFG (German Research Foundation) – Project-ID 273811115 – SFB 1225 ISOQUANT (SJ), the Germany’s Excellence Strategy EXC2181/1-390900948 (Heidelberg Excellence Cluster STRUCTURES) (SJ) and the European Union’s Horizon 2020 research and innovation program under grant agreement No.~817482 (PASQuanS) (SJ), grants from the Knut and Alice Wallenberg Foundation (Grant No. KAW~2023.0322) (SMR) and the Swedish Research Council (Grant No. 2022-03654 VR) (SMR). This work has been partially financed by the Baden-Württemberg Stiftung. SB acknowledges funding by the Harvard Quantum Initiative.

\paragraph*{Author Contributions}
S.B. and C.H. performed the measurements. S.B. analyzed the data.  S.B. calculated the mean-field BCS theory LDA curves. G.B. computed the correlators in the dimer limit. S.B. and G.B. computed the correlators in the potential dominated regime. F.B.  performed the exact diagonalization calculations with conceptual input from S.M.R. . M.G. and S.J. supervised the experimental part of the project. S.B. wrote the manuscript with input from all authors. All authors contributed to the discussion of the results.

\input{methods}
\end{document}

%% file: methods.tex
\title{Supplemental material: Cooper pairs --  from the surface to the bulk}

\date{\today}


\maketitle


\setcounter{figure}{0}
\renewcommand{\figurename}{Extended Data Figure}
\renewcommand{\theequation}{S\arabic{equation}}

\section{\bf Methods.}
\subsection{Experimental details}

We start the experimental sequence by laser cooling $^6$Li atoms using a Zeeman slower and a magneto optical trap. From there the atoms are transferred into a red detuned, crossed beam optical dipole trap, where we perform a sequence of radio frequency pulses to obtain a balanced mixture in hyperfine states $\ket{1}$ and $\ket{3}$ of the $^2$S$_\frac{1}{2}$ Lithium ground state manifold (with states $\ket{1}-\ket{6}$ labeled in ascending order of energies). After a short evaporation, the atoms are transferred into a tightly focused optical tweezer, created using light with a wavelength of \SI{1064}{\nano \meter}. Making use of the high densities we perform fast evaporation in this optical tweezer, followed by the spilling technique described in~\cite{Serwane_2011} to arrive at $\approx 30$ atoms. Subsequently we perform a continuous crossover to the quasi-2D regime by ramping on the power of a vertical optical lattice, created by two beams with a wavelength of $\SI{1064}{\nano \meter}$, interfering under an angle,  creating a light-sheet. This light sheet provides a strong vertical confinement with a trap frequency of trap frequency $\omega_z/2\pi = \SI{7423(3)}{\hertz}$. Simultaneously we weaken the radial confinement of the optical tweezer, resulting in a radial trap frequency on the order of $\omega_r/2\pi \approx \SI{1000}{\hertz}$. To ensure that the system remains quasi-2D, both $E_\text{F}$ and $E_\text{B}$ need to be smaller than $\hbar\omega_\text{z}$. This is ensured by choosing a lower radial trap frequency $\omega_\text{r}$ in the limit of high $E_\text{F}$ and $E_\text{B}$. The low atom number before this transfer allows us to ensure that atoms are only loaded into single layer of the vertical optical lattice  (2D-OT). Using the spilling technique introduced in~\cite{Bayha2020}, we deterministically prepare different atom numbers in the ground state of this 2D optical tweezer. A more detailed account of this preparation sequence can be found in~\cite{Bayha2020}.

We obtain the correlator \(C^{(2)}(r,r)\) from the measured positions by forming all opposite-spin pairs in each experimental realization. For atoms at positions \(\mathbf{r}_{i,\uparrow}\) and \(\mathbf{r}_{j,\downarrow}\), we define
\[
\mathbf{R}_{ij}=\frac{\mathbf{r}_{i,\uparrow}+\mathbf{r}_{j,\downarrow}}{2},
\qquad
\mathbf{s}_{ij}=\mathbf{r}_{i,\uparrow}-\mathbf{r}_{j,\downarrow}.
\]
We retain pairs with \( |\mathbf{s}_{ij}|<\epsilon \), where $\epsilon$ is set by the experimental resolution and bin their centre-of-mass radii \(R_{ij}=|\mathbf{R}_{ij}|\). For a radial bin \([r_k,r_{k+1})\), we estimate
\[
C^{(2)}_{\rm exp}(r_k,r_k)
=
\frac{
N^2\left(N_k-N_k^{\rm rand}\right)
}{
N_{\rm shots}\,
N_{\rm pair}\,
\pi(r_{k+1}^2-r_k^2)\,
\pi\epsilon^2
}.
\]
Here \(N_k\) is the number of accepted pairs accumulated over all analyzed realizations, \(N_k^{\rm rand}\) is the corresponding randomized reference count (obtained from correlations between spin up and down atoms in different experimental realizations), \(N_{\rm shots}\) is the number of experimental realizations, and \(N_{\rm pair}\) is the total number of opposite-spin pairs entering the analysis. The factor \(N^2\), with \(N\) the atom number per spin state, normalizes the signal to the number of possible opposite-spin pairs, such that a fully paired sample corresponds to \(N^2\) pairs per realization. The factors \(\pi(r_{k+1}^2-r_k^2)\) and \(\pi\epsilon^2\) are the centre-of-mass annulus area and the accepted relative-coordinate area, respectively. We obtain the momentum-space correlator \(C^{(2)}(p,-p)\) analogously, replacing the real-space coordinates by the measured momenta and selecting pairs with small total momentum, i.e. \(|\mathbf{p}_{i,\uparrow}+\mathbf{p}_{j,\downarrow}|<\epsilon_p\), while binning the corresponding relative momentum.

\subsubsection{Quasi-2D model for the tweezer trap}

Due to the strong vertical confinement $\omega_z$, which is more than seven times larger than the radial confinement $\omega_r$ of the optical tweezer, one may separate out the $z$-direction,  $V(x,y,z)=V_{\mathrm{2D}}(x,y) + V_z(z)$,  
leading to the effective 2D Hamiltonian 
\begin{equation}
    \hat H_{2D}=\sum _{i=1}^{2N}\left[-\frac{\hbar ^2}{2m_a} \nabla _i^2 + 
    V_{\mathrm{2D}}({\bf r} _i) \right] + g\sum _{k,l}\delta (\hat{{\bf r}}_k-\hat{{\bf r}}_l)
    \label{eq:Nfermion_Hamiltonian}
\end{equation}
for $2N$ fermions in two different hyperfine (pseudospin) states $(k,l)$, 
with ${\bf r}_i=(x_i,y_i)$ and $\nabla _i =\partial _{x_i} +\partial _{y_i}$. (Note that contact interactions do not couple fermions with the same pseudospin.) 
In the limit of small atom numbers, earlier experiments have shown that the tweezer trap can be modeled by a  slightly asymmetric and anharmonic Gaussian profile. For the details we refer to the earlier work by Bayha {\it et al.}~\cite{Bayha_2020}.  
To the lowest order, the confinement is harmonic, with 
single-particle eigenstates  
{\small
\begin{equation}
	\braket{{\bf r}|n,m} = \sqrt{\frac{k!}{\pi (k+|m|)!}} (\frac{r}{l_\text{HO}})^{|m|}e^{-\frac{1}{2}{(r/l_\text{HO})^2}} L_k^{|m|}(r^2)e^{im\theta}
	\label{eq: pairing_HO_wavefunction}
\end{equation}}
in real space coordinates $(x,y)=(r,\theta)$ with harmonic oscillator length $l_\text{HO}=\sqrt{\hbar/m_a\omega_\text{r}}$. 
Here, $n$ is the principal/shell quantum number, $m=-n,-n+2,..+n$ is the angular momentum quantum number, and $L_k^{|m|}$ is the generalized Laguerre polynomial of degree $k$, with 
$k = \frac{1}{2}(n-|m|)$.   
The single-particle eigenenergies are 
$E^0_n = (n+1)\hbar \omega_\text{r}$ with a $n+1$-fold degeneracy.

\subsection{Exact diagonalization}
\begin{figure}
    \includegraphics{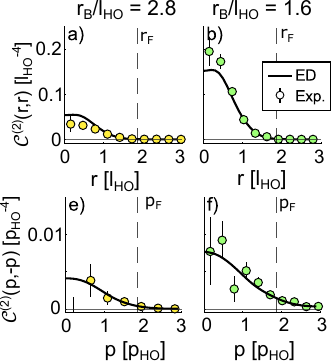}
    \caption{\textbf{Comparison to exact diagonalization data.} 
    Shown are the density-density correlators $\mathscr{C}^{(2)}(r,r)$ in real space (upper panels) at the same position 
    ${\bf r}_\uparrow ={\bf r}_\downarrow $ at a given radius 
    $r=\mid {\bf r}_{\sigma }\mid $ 
    and $\mathscr{C}^{(2)}(p, -p)$ in momentum space (lower panel), for $N+N=1+1$. Units for position and momentum are  oscillator length
    $l_{\mathrm{HO}}=\sqrt{\hbar/m_a\omega _r}$ and momentum $p_{\mathrm{HO}}=\sqrt{\hbar m_a\omega _r}$.}
    \label{fig: Comparison ED 1+1}
\end{figure}

\begin{figure}
    \includegraphics{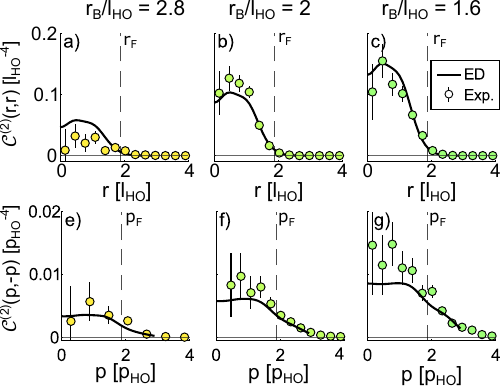}
    \caption{\textbf{Comparison to exact diagonalization.} As in Fig.~\ref{fig: Comparison ED 1+1} but for $N+N=3+3$.}
    \label{fig: Comparison ED N+N=3+3}
\end{figure}

In the few-body limit 
the effective 2D Hamiltonian Eq.~\eqref{eq:Nfermion_Hamiltonian} can be directly diagonalized by applying the configuration interaction approach, as detailed in Refs.~\cite{Bjerlin_2016,Bayha_2020}. Here, we follow the numerical methodology of Ref.~\cite{Bayha_2020}, using the same parameters for the trap when comparing densities and correlation functions with the experimental data for up to 3+3 fermions. As explained in Refs.~\cite{Bjerlin_2016,Rontani_2009,Bayha_2020}, the contact interaction in a given many-body Hilbert space is re-normalized via comparison with the exact two-body binding energy.  The single-particle energy cutoff used here was $10\hbar \omega _r $ and the non-interacting many-body energy cutoff was $28\hbar \omega _r $.  This defines the subspace of many-body Fock states constructed from the single-particle oscillator states $\braket{{\bf r}|n,m}$. The real-space and momentum-space density-density correlators $\mathscr{C}^{(2)}(\mathbf{r}_\uparrow,\mathbf{r}_\downarrow )$ and $\mathscr{C}^{(2)}(\mathbf{p}_\uparrow,\mathbf{p}_\downarrow )$, as defined in the main text,  are calculated  from the corresponding density operators ${\hat n}({\bf r}_{\sigma })$ and ${\hat n}({\bf p}_{\sigma })$ with $\sigma \in \{ {\uparrow,\downarrow}\}$, where  the respective expectation values are determined using the corresponding numerically exact many-body eigenstates. The computed correlators in both real and momentum space are in excellent agreement with the experimental measurements, as shown in Figures \ref{fig: Comparison ED 1+1} and \ref{fig: Comparison ED N+N=3+3}, for systems of $1+1$ and $3+3$ atoms, respectively. This agreement provides an important benchmark for our analysis, validating that the experimentally extracted correlators faithfully capture the underlying few-body correlations. In the potential-dominated regime for larger atom numbers, the numerical effort of exact diagonalization becomes unfeasible. 


\subsubsection{Shell pairing}
In the confinement dominated regime where $r_B\gtrsim r_F$, the pairing mainly occurs between time-reversed states within the same 
2D harmonic oscillator shell, i.e.\ between single particle states $\ket{n,m,\uparrow}$ and $\ket{n,-m,\downarrow}$. 
We may then simplify the Hamiltonian to a BCS model describing the pairing between these time-reversed states~\cite{Anderson1959,Bruun_2014}
\begin{equation}
\begin{split}
    	\hat{H} & = \sum_{n,m,\sigma} \epsilon_n \hat{a}^\dagger_{n,m,\sigma} \hat{a}_{n,m,\sigma} \\
        & + \sum_{\substack{n,n',m,m'}} V_{n,m,n',m'}\hat{a}^\dagger_{n,m,\uparrow} \hat{a}^\dagger_{n,-m,\downarrow} \hat{a}_{n',-m',\downarrow} \hat{a}_{n',m',\uparrow}
\end{split}
\end{equation}
where $\epsilon_n = (n+1)\hbar \omega_\text{r} - E_\text{F}$ denotes the energy of state $n$ relative to the Fermi energy and $\hat{a}^\dagger_{n,m,\sigma}$ $(\hat{a}_{n,m,\sigma})$ are the creation(annihilation) operators of a particle  in state $\ket{n,m,\sigma}$. Employing BCS mean-field theory, the Hamiltonian can be rewritten as 
\begin{equation}
	\hat{H} = \sum_{n,m,\sigma} \epsilon_n \hat{a}^\dagger_{n,m,\sigma} \hat{a}_{n,m,\sigma} + \sum_{n,m} \Delta (\hat{a}^\dagger_{n,m,\uparrow} \hat{a}^\dagger_{n,-m,\downarrow} + \text{h.c.}),
\end{equation}
where $\Delta$ is the pairing gap, which is determined using a self-consistent gap equation~\cite{Bruun_2014}.

This Hamiltonian can be diagonalized using the Bogoliubov de Gennes transformation~\cite{Bogoljubov_1958}, i.e. by introducing quasi particle operators
\begin{equation}
	\begin{split}
		\hat{\gamma}_{n,m, \uparrow} = u_{n,m} \hat{a}_{n,m,\uparrow} -v^{*}_{n,-m} \hat{a}^\dagger_{n,-m,\downarrow} \\
		\hat{\gamma}^\dagger_{n,m, \downarrow} = u^{*}_{n,-m} \hat{a}_{n,-m,\downarrow} +v_{n,m} \hat{a}^\dagger_{n,m,\uparrow},
	\end{split}	
\end{equation}
with $u^2_{n,m} = 1/2(1+\epsilon_n/E_n) = 1 -v^2_{n,m}$ and $E_n = \sqrt{\epsilon_n^2+\Delta^2}$. 

The pairing gap $\Delta$ depends on the occupation of the last filled harmonic oscillator shell $n_\text{F}$ ~\cite{Heiselberg_2002,Bruun_2014}. When there are unoccupied states in the last shell ("open shell"), the Fermi energy is equal to the energy of this shell $E_\text{F} = (n_\text{F}+1)\hbar \omega_\text{r}$. Here, pairing will occur at arbitrarily weak interactions -- similarly to the continuous system -- within the last filled shell $n_\text{F}$. However, when all states up to and including $n_\text{F}$ are completely filled, the Fermi energy lies between the last filled and the first empty shell $E_\text{F} = (n_\text{F}+3/2)\hbar \omega_\text{r}$. The gap in the single-particle spectrum (set by $\hbar \omega_\text{r}$) 
then gives rise to a critical interaction strength, at which the system transitions from an unpaired to a paired state. At this critical interaction strength, it becomes energetically favorable to excite two atoms from the Fermi surface to the first unfilled shell, thus permitting the formation of pairs.  The emergence of this phase transition was predicted in~\cite{Bruun_2014} and observed in~\cite{Bayha_2020}.

The critical binding energy $E_\text{B}^\text{c}$ in the filled shell configuration, is given by~\cite{Bruun_2014}
\begin{equation}
	E_\text{B}^\text{c}/\hbar \omega_\text{r} = \frac{B(n_\text{F})}{\zeta(2)} \left(\sqrt{1+4\frac{\zeta(2)}{B(n_\text{F})^2}}-1\right),
\end{equation}
where $B(n_\text{F}) = 0.577+4\ln 2 + \ln n_\text{F}$ and $\zeta$ is the Riemann zeta function. Below this critical binding energy, pairing is energetically unfavorable. For binding energies $E_\text{B}$ larger than the critical binding energy, the pairing gap is given by~\cite{Bruun_2014}
\begin{equation}
	\Delta/\hbar \omega_\text{r} = \frac{1}{\sqrt{7\zeta(3)}} \sqrt{\frac{2 \hbar \omega_\text{r}}{E_\text{B}^\text{c}}-\frac{2 \hbar \omega_\text{r}}{E_\text{B}}+\zeta(2)\left(\frac{E_\text{B}}{2 \hbar \omega_\text{r}}-\frac{E^\text{c}_\text{B}}{2 \hbar \omega_\text{r}}\right)}.
	\label{eq:pairing_Delta_potential_dominated}
\end{equation}
 Note that these results were derived in the mean-field limit. In the few-body limit the phase transition is softened to a crossover, and pairing can also occur below the critical interaction strength~\cite{Bayha_2020,Bjerlin_2016}.

\subsection{Local density approximation BCS limit}\label{Sec:BCStheory}
In the limit where the system can be treated as locally homogeneous, the introduction of a local Fermi energy is justified, replacing the global Fermi energy in the potential-dominated regime. The Fermi energy is set by the particle density $n_\sigma(r)$ of a single spin component,
\begin{equation}
	E_\text{F}(r) = \frac{\hbar^2}{2m_a} k_\text{F}(r)^2 = \frac{2\pi \hbar^2}{m_a} n_\sigma(r).
\end{equation}
This allows us to define a local pairing gap $\Delta(r) = \sqrt{2E_\text{F}(r)E_\text{B}}$, and this approximation is commonly known as the local density approximation (LDA).

The momentum space correlator is for a homogeneous system
\begin{align}
	\mathscr{C}^{(2)}(p,-p) &=
    \underbrace{\braket{\hat{a}^\dagger_{p\uparrow}\hat{a}_{p\uparrow}\hat{a}^\dagger_{-p\downarrow}\hat{a}_{-p\downarrow}}}_{\braket{\hat{n}_\uparrow(p)\hat{n}_\downarrow(-p)}}-\underbrace{\braket{\hat{a}^\dagger_{p\uparrow}\hat{a}_{p\uparrow}}\braket{\hat{a}^\dagger_{-p\downarrow}\hat{a}_{-p\downarrow}}}_{\braket{\hat{n}_\uparrow(p)}\braket{\hat{n}_\downarrow(-p)}} \nonumber\\
    &= (v_p u_p)^2 = \frac{\Delta^2}{4(\epsilon_p^2 + \Delta^2)}.
	\label{eq:pairing_cooper_MS}
\end{align}
Here, $\braket{...} \equiv \braket{\Psi_\text{BCS}|...|\Psi_\text{BCS}}$ is the ground state expectation value and $v_p^2 = \frac{1}{2} \left(1- \frac{\epsilon_p}{E_p}\right)$ and $u_p^2 = \frac{1}{2} \left(1+ \frac{\epsilon_p}{E_p}\right)$ are the usual coherence factors
with $E_p = \sqrt{\epsilon_p^2+|\Delta|^2}$. The energy $\epsilon_p = \frac{p^2}{2m}-\mu$ is that of the state with momentum $p$, relative to the chemical potential $\mu$. 
In a trap, we have to consider that the momentum space measurement averages over all positions, i.e. over regions of different densities, resulting in locally varying Fermi energies.  The momentum space correlator then becomes
\begin{equation}
\begin{split}
\mathscr{C}^{(2)}(p,-p)
&=
\left(
\int_{0}^{r_\text{max}} dr\,
\frac{r}{2\pi\hbar^2}
\right. \\
&\qquad\left.
\times
\frac{\Delta(r)}
{2\sqrt{\left(\frac{p^2}{2m_a}-\mu(r)\right)^2+\Delta(r)^2}}
\right)^2 .
\end{split}
\label{eq:LDA_MS}
\end{equation}

The real-space correlation function in Eq.~\eqref{eq: pairing_general_corr} is likewise given by 
\begin{equation}
  \mathscr{C}^{(2)}(\mathbf{r}_\uparrow,\mathbf{r}_\downarrow) = |\braket{\hat\psi_\uparrow(\mathbf{r}_\uparrow)\hat\psi_\downarrow(\mathbf{r}_\downarrow)}|^2 
	\label{eq: pairing_general_corrBCS}
\end{equation}
where 
\begin{equation}
\braket{\hat\psi_\uparrow(\mathbf{r}_\uparrow)\hat\psi_\downarrow(\mathbf{r}_\downarrow)}=\frac\Delta2\int\!\frac{d^2p}{(2\pi)^2}
\frac{e^{i\mathbf{p}\cdot({\mathbf r}_\uparrow-{\mathbf r}_\downarrow)}}{E_p}
\label{LogInt}
\end{equation}
and $\hat\psi_\sigma(\mathbf{r})$ removes a spin $\sigma$ particle at position $\mathbf{r}$.
It is easy to show that the integral in Eq.~\eqref{LogInt} is logarithmically divergent so that 
\begin{equation}
\lim_{|{\mathbf r}_\uparrow-{\mathbf r}_\downarrow|\rightarrow 0}\braket{\hat\psi_\uparrow(\mathbf{r}_\uparrow)\hat\psi_\downarrow(\mathbf{r}_\downarrow)}=-\frac{\Delta m}{2\pi}\ln(k_F|{\mathbf r}_\uparrow-{\mathbf r}_\downarrow|).
\label{LogDivergence}
\end{equation}
This divergence comes from assuming a short range interaction. At very small relative distances between particles, the physics is governed by the two-body contact, which changes the prefactor of the logarithmic divergence from the BCS prediction 
in Eq.~\eqref{LogDivergence}~\cite{Werner2012}. 

Experimentally, the finite resolution of the matter wave magnifier introduces an effective cutoff, as atoms with very small relative position, i.e. high relative momenta, are not accurately magnified. 
Finite range effects will also remove this logarithmic divergence.

When the pairing gap varies locally, the real space pair density becomes spatially dependent with
\begin{equation}
	\mathscr{C}^{(2)}(r,r) = \left(\int_{p=0}^{p_\text{max}} dp \frac{p} {2 \pi \hbar^2} \frac{\Delta(r)}{2\sqrt{\left(\frac{p^2}{2m_a}-\mu(r)\right)^2+\Delta(r)^2}}\right)^2,
	\label{eq: LDA_RS}
\end{equation}
where we have introduced an ultraviolet cut-off $p_\text{max}$ to describe  the effects of 
a finite interaction range  and the experimental resolution discussed above.

 \subsection{Dimer limit}

In the BEC limit $r_B\ll \ell$, the system becomes a gas of tightly bound dimers. For a homogenuous system of area $\mathcal A$, the wave function for $N$ such dimers is then  accurately approximated by 
\begin{equation}
\ket{\Psi}=\frac1{\sqrt{N!}}\left(\frac 1{\sqrt\mathcal A}\int\!d^2\!R\,\hat\Phi^\dagger({\mathbf R})\right)^N\ket{0}
\end{equation}
where 
\begin{equation}
\hat\Phi^\dagger({\mathbf R})=\int\!d^2r\,\phi({\mathbf r})\hat\psi^\dagger_\uparrow({\mathbf R}+{\mathbf r}/2)\hat\psi^\dagger_\downarrow({\mathbf R}-{\mathbf r}/2)
\end{equation}
creates a dimer with  wave function $\phi({\mathbf r})$ at center-of-mass position ${\mathbf R}$. Using this wave function, it is then easy to show that 
$\mathscr{C}^{(2)}(\mathbf{r}_\uparrow,\mathbf{r}_\downarrow)\rightarrow |\phi({\textbf r})|^2n/2$ for 
$|\mathbf{r}_\uparrow-\mathbf{r}_\downarrow|=r\rightarrow 0$ with $n=2N/\mathcal A$ the total density of atoms, i.e.\ twice the density of dimer. Within the local density approximation, we then simply use the local total density $n\rightarrow n({\mathbf r})$.

Since $\phi({\mathbf r})\sim \ln(r/r_B)$ for $r\rightarrow0$ for fermions interacting via short range interaction, $\mathscr{C}^{(2)}(r,r)$ is strictly speaking logarithmically divergent~\cite{Werner2012}. This 
divergence is the same as we encountered in the BCS calculation above and it is effectively integrated out by our experimental resolution and by finite range effects. We therefore compare the experimental data in Fig.~\ref{fig: Interactions}c with the function $\alpha n(r)$ where $\alpha$ is a fit parameter and $n(r)$ is the measured density of the system.
